\documentclass{article}
\usepackage{dcase2020,amsmath,graphicx,url,times,booktabs, tabularx}
\usepackage[table,xcdraw]{xcolor}

\def \am [#1]{\textcolor{red}{AM: #1}}

\title{Acoustic scene classification in DCASE 2020 Challenge: \\ generalization across devices and low complexity solutions}
\name{Toni Heittola, Annamaria Mesaros, Tuomas Virtanen}
 \address{Computing Sciences \\ Tampere University, Finland\\
 \{toni.heittola, annamaria.mesaros, tuomas.virtanen\}@tuni.fi}
\begin{document}

\ninept
\maketitle

\begin{sloppy}

\begin{abstract}
This paper presents the details of Task 1 Acoustic Scene Classification in the DCASE 2020 Challenge. The task consisted of two subtasks: classification of data from multiple devices, requiring good generalization properties, and classification using low-complexity solutions. Each subtask received  around 90 submissions, and most of them outperformed the baseline system. The most used techniques among the submissions were data augmentation in Subtask A, to compensate for the device mismatch, and post-training quantization of neural network weights in Subtask B, to bring the model size under the required limit. The highest classification accuracy on the evaluation set in Subtask A was 76.5\%, compared to the baseline performance of 51.4\%. 
In Subtask B, many systems were just below the 500~KB size limit, and the maximum classification accuracy was 96.5\%, compared to the baseline performance of 89.5\%. 
\end{abstract}


\section{Introduction}
\label{sec:intro}

The goal of acoustic scene classification is to classify a test audio recording into one of the provided predefined classes that characterizes the environment in which it was recorded. Acoustic scene classification is a popular task of the DCASE Challenge, and brings new variations of a supervised classification task each year. In all previous editions, this task has attracted the highest number of participants among the available tasks.

Recent years have seen a boom in deep-learning based solutions for various classification problems \cite{Nguyen2018a, Koutini2019}, also obvious in the last few editions of the DCASE Challenge. While in 2016 just 22 of the 48 submissions used neural networks \cite{Mesaros2018_TASLP} and many other methods had top performance \cite{Bisot2016}, in 2019 only five of the 146 systems submitted to the acoustic scene classification subtasks did not include a deep learning component \cite{mesaros_2019_DCASE}. Generally, deep learning algorithms require large amounts of data for best performance, and the effort to produce more data for the task has resulted in gradual extension of the problem, from a classical textbook example \cite{Mesaros2018_TASLP, Lidy2016,  Mesaros2018a} to domain adaptation due to mismatched devices \cite{Mesaros2018_DCASE, Primus2019}, and open-set classification \cite{mesaros_2019_DCASE, Wilkinghoff2019}. 

\begin{figure}
    \centering
    \includegraphics[width=0.95\columnwidth]{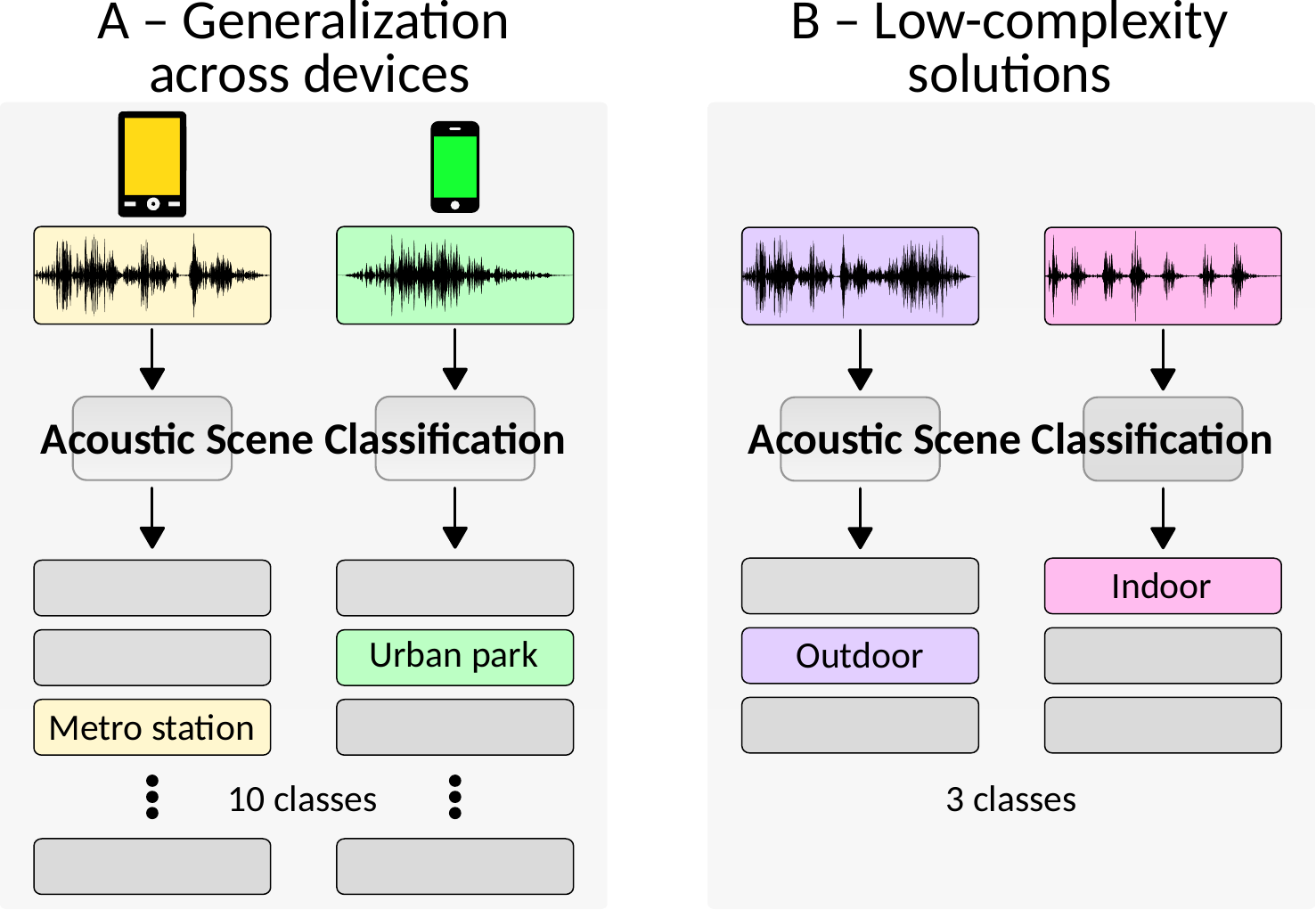}
    \vspace{-6pt}
    \caption{Acoustic scene classification in DCASE 2020 Challenge. Overview of the two tasks.}
    \label{fig:subtaskA}
\end{figure}

Deep learning solutions are sensitive to mismatch between training and testing data, which is often present in realistic scenarios because of the large variety of devices available for recording audio. While in previous editions of the challenge the device mismatch was treated from the point of view of domain adaptation and targeted for a small number of test devices, it is realistic to assume that any method tested in a real-world scenario would have to face a much higher number of devices, many of which are not available at training stage. 
Furthermore, because the main application area of acoustic scene classification is context-aware devices, where the algorithms should be running on devices with limited computational capacity, taking into account computational limitations is another important point for real-world applications. The acoustic scene classification task in DCASE 2020 Challenge brings these two topics into the spotlight.

In DCASE 2020 Challenge, the acoustic scene classification task comprises two different subtasks, which require system development for two different situations. 
Subtask A: Acoustic Scene Classification with Multiple Devices requires classification of data from real and simulated devices, and is aimed for developing systems with very good generalization properties across a large number of different devices. For this subtask, the challenge provides new data consisting of real and simulated audio recordings for mobile devices.  
Subtask B: Low-Complexity Acoustic Scene Classification requires classification of data from a single device, and is aimed for developing low-complexity solutions for the problem. In this subtask, the challenge imposes a maximum model size as a proxy for estimating the model complexity at test time. 

This paper introduces the two acoustic scene classification tasks and their results and is organized as follows: Sections~\ref{sec:subtaskA} and~\ref{sec:subtaskB} introduce the setup, dataset and baseline system for subtasks A and B, respectively. Section~\ref{sec:results} presents the challenge results for both subtasks, and an analysis of selected submissions. Finally, Section~\ref{sec:conclusions} presents conclusions and future perspectives on this task for upcoming editions of the challenge.

\section{Acoustic Scene Classification with Multiple Devices}
\label{sec:subtaskA}

This subtask is concerned with the basic problem of acoustic scene classification, in which it is required to classify a test audio recording into one of ten known acoustic scene classes. A specific feature for this edition is generalization across a number of different devices, enforced through use of audio data recorded and simulated with a variety of devices. 

\subsection{Dataset and performance evaluation}

as explained A new dataset was created for this task, called \textbf{TAU Urban Acoustic Scenes 2020 Mobile} \cite{Dataset2020_Mobile_dev,Dataset2020_Mobile_eval}. The dataset is based on the TAU Urban Acoustic Scenes 2019 dataset, containing recordings from multiple European cities and ten different acoustic scenes \cite{mesaros_2019_DCASE}. The new dataset contains the four devices used to record simultaneously (A, B, C, and D), and additional synthetic devices S1-S11 simulated using audio recorded with device A which is a high quality binaural device.

Simulated recordings were created using a dataset of impulse responses (IR) measured for multiple angles using mobile devices other than the ones already in the dataset (B, C, D). To simulate the recording from a device S, audio recorded with device A was processed through convolution with the device-specific IR, followed by a dynamic range compression with device-specific parameters. The IR angle was randomly selected among the available ones, and is location-specific, to simulate the case when a long recording has been captured with device S in a certain position. 

The development set contains data from ten cities and nine devices: three real devices (A, B, C) and six simulated devices (S1-S6). Data from devices B, C and S1-S6 consists of randomly selected segments from the simultaneous recordings, therefore all overlap with the data from device A, but not necessarily with each other. The total amount of audio in the development set is 64 hours, of which 40 hours are from device A. Audio was provided in single channel 44.1~kHz 24-bit format. The dataset was provided with a training/test split in which 70\% of the data for each device is included for training, 30\% for testing; some devices appear only in the test subset. Complete details are presented in Table 1. 

The evaluation dataset contains 33 hours of audio from all 12 cities, ten acoustic scenes, 11 devices. Five of the 11 devices from the evaluation set are unseen in training (not available in the development set): real device D and simulated devices S7-S11. 
Device and city information is not provided in the evaluation set, as the systems are expected to be robust to different devices. 

Evaluation of submissions was performed using two metrics: accuracy and multi-class cross-entropy. Accuracy was calculated as macro-average (average of the class-wise accuracy for the acoustic scene classes). Multi-class cross-entropy (log loss) was used in order to have a metric which is independent of the operating point. Ranking of the systems was be done by accuracy. 

\subsection{Baseline system results}

The baseline system  provided for the task uses Open L3 embeddings as feature representation, followed by
two fully-connected feed-forward layers \cite{Cramer_2019_ICASSP}.
The system uses a window size of one second for analysis, with a hop size of 100 ms, input representation through 256 mel filters, content type music, and an embedding size of 512. This is followed by two fully connected layers of 512 and 128 hidden units, respectively. The learning is performed for 200 epochs with a batch size of 64,  and data shuffling between epochs, using Adam optimizer \cite{Kingma2015} with a learning rate 0.001. Model selection is performed using validation data consisting of approximately 30\% of the original training data.
Model performance after each epoch is evaluated on the validation set, and the best performing model is selected. 

The overall performance of the baseline system on the development set is 54.1\%. The baseline system does not have any mechanism for explicitly dealing with the device mismatch, which is evident from the results: device-wise system performance decreases according to the amount of data available in training. Highest accuracy is obtained on device A (70.6\%), while the other devices for which a small amount of data is available in training provide about 50-60\% accuracy; the lowest accuracy is observed for the unseen devices (39-48\%).

\section{Low-Complexity Acoustic Scene Classification}
\label{sec:subtaskB}

\subsection{Description}
This subtask is concerned with classification of audio into three major acoustic scene classes, with focus on low complexity solutions for the classification problem in term of model size, and uses audio recorded with a single device (device A).

\begin{table} 
\label{tab:subtaskA-data}
\centering
\caption{Distribution of devices among training and test subsets in the dataset. Some devices are present only in the test set to simulate real situations of encountering an unseen device at the usage end.}
\setlength\tabcolsep{4.3pt}
\begin{tabular}{ll|ll|lll}
\toprule  
\multicolumn{2}{c|}{\textbf{Device}} &
  \multicolumn{2}{c|}{\textbf{Dataset}} &
  \multicolumn{3}{c}{\textbf{Cross-validation setup}} \\
{\textbf{Name}} &
  {\textbf{Type}} &
  {\textbf{\begin{tabular}[c]{@{}l@{}}Total\\ length\end{tabular}}} &
  {\textbf{\begin{tabular}[c]{@{}l@{}}Total\\ seg\end{tabular}}} &
  {\textbf{\begin{tabular}[c]{@{}l@{}}Train\\ seg\end{tabular}}} &
  {\textbf{\begin{tabular}[c]{@{}l@{}}Test\\ seg\end{tabular}}} &
  {\textbf{\begin{tabular}[c]{@{}l@{}}Unused\\ seg\end{tabular}}} 
  \\
\midrule  
\cellcolor[HTML]{a3d7a3}A &
\cellcolor[HTML]{a3d7a3}Real &
  {40h} &
  {14400} &
  {10215} &
  {330} &
  {3855}
  \\
\cellcolor[HTML]{a3d7a3}B,C &
\cellcolor[HTML]{a3d7a3}Real &
  {3h*2} &
  {1080*2} &
  {750*2} &
  {330*2} &
  \\
\cellcolor[HTML]{f4c884}S1,S2,S3 &
\cellcolor[HTML]{f4c884}Sim. &
  {3h*3} &
  {1080*2} &
  {750*3} &
  {330*3} &
  \\
\cellcolor[HTML]{f4c884}S4,S5,S6 &
\cellcolor[HTML]{f4c884}Sim. &
  {3h*3} &
  {1080*3} &
  - &
  {330*3} &
  {750*3}
 \\
\midrule
{\textbf{Total}} &
{} &
{\textbf{64h}} &
{\textbf{23040}} &
{\textbf{13965}} &
{\textbf{2970}} &
{\textbf{6105}} \\
\bottomrule
\end{tabular}
\end{table}

\begin{table*}[]
\centering
\caption{Model size for the two subtask baselines and their performance on Subtask B.}
\label{tab:model-size}
\begin{tabular}{
>{}l l
>{}c 
>{}c 
>{}c 
>{}l 
>{}l }
\toprule
{\textbf{System}} &
{\textbf{Implementation details}} &
{\textbf{Accuracy}} &
{\textbf{Log loss}} &
{\textbf{Audio embeddings}} &
{\textbf{Ac. model}} &
{\textbf{Total size}} \\
\midrule
\textbf{Subtask B Baseline} &
 \begin{tabular}[c]{@{}l@{}}Log mel-band energies + CNN \\ (2 CNN layers + 1 fully-connected)\end{tabular} &
 \begin{tabular}[c]{@{}l@{}}87.3 \% \\ ($\pm$ 0.7)\end{tabular} &
 \begin{tabular}[c]{@{}l@{}}0.43 \\ ($\pm$ 0.04)\end{tabular} &
 - &
 450.1 KB &
 450.1 KB \\
 {Subtask A Baseline} &
  \begin{tabular}[c]{@{}l@{}}OpenL3 + MLP \\ (2 layers, 512 and 128 units)\end{tabular} &
  {\begin{tabular}[c]{@{}l@{}}89.8 \% \\ ($\pm$ 0.3)\end{tabular}} &
  {\begin{tabular}[c]{@{}l@{}}0.26 \\ ($\pm$ 0.00)\end{tabular}} &
  17.87 MB &
  145.2 KB &
  19.12 MB \\
 \bottomrule
 \end{tabular}
\end{table*}

\begin{table*}[]
\centering
\caption{Selected systems submitted for Subtask A. Top 10 teams, best system per team.}
\label{tab:subtaskA-top10}
\setlength\tabcolsep{3.2pt}
\begin{tabular}{ll|cc|cc|l}
\toprule
{\textbf{System}} & {\textbf{\#}} & {\textbf{Accuracy}}   & {\textbf{Log loss}}  &  {\textbf{Seen devices}} & {\textbf{Unseen devices}} & {\textbf{System summary, \textit{approach for device mismatch}}}    \\
\midrule
Suh\_ETRI\_3        & 1 & {76.5 \%}    & {1.21}      & {78.1 \%} & {74.6 \%} & {ResNets} \\ 
Hu\_GT\_3         & 3 & {76.2 \%}    & {0.89}      & {77.5 \%} & {74.7 \%} & Two-stage classification, \textit{data augmentation} \\ 
Gao\_UNISA\_4     & 8    & {75.2 \%}    & {1.23}      & {77.0 \%} & {73.1 \%} & ResNets, \textit{domain adaptation} \\
Jie\_Maxvision\_1  & 10  & {75.0 \%}    & {1.20}      & {76.5 \%} & {73.2 \%} & ResNet with attention mechanism (single model)\\
Koutini\_CPJKU\_3  & 13 & {73.6 \%}    & {1.20}      & {76.8 \%} & {69.8 \%} & ResNets, \textit{domain adaptation}, Rf reg. \\
Helin\_ADSPLAB\_1  & 14 & {73.4 \%}    & {0.85}      & {76.2 \%} & {70.1 \%} & CNNs (pretrained), multiple decision schemes \\
Liu\_UESTC\_1      & 16 & {73.2 \%}    & {1.30}      & {74.3 \%} & {71.9 \%} & ResNets using spectrogram decompositions \\ 
Zhang\_THUEE\_2     & 17 & {73.2 \%}    & {1.96}     & {75.8 \%} & {70.0 \%} & ResNets and SegNets, \textit{spectrum correction} \\
Lee\_CAU\_4         & 20 & {72.9 \%}    & {0.91}    & {75.5 \%} & {69.8 \%} & CNNs \\
Liu\_SHNU\_4        & 26 & {72.0 \%}    & {3.16}    & {75.8 \%} & {67.5 \%} & ResNet and CNN, \textit{data augmentation} \\
\midrule
{Baseline}          &        & {51.4 \%}    & {1.90}   & {63.1 \%} & {37.2 \%} & OpenL3, single MLP model \\
\bottomrule
\end{tabular}
\end{table*}

\subsection{Dataset and performance evaluation}

A new dataset was created for this task, called \textbf{TAU Urban Acoustic Scenes 2020 3Class} \cite{Dataset2020_3Class_dev,Dataset2020_3Class_eval}, based on TAU Urban Acoustic Scenes 2019. The ten acoustic scenes from the original data were grouped into three higher level acoustic scene classes as follows: indoor (airport, indoor shopping mall, and metro station), outdoor (pedestrian street, public square, street with medium level of traffic, and urban park), and transportation (travelling by bus, travelling by tram, and travelling by underground metro). This dataset contains audio recorded with a single device (device A),  provided in binaural, 48 kHz 24-bit format.
The development dataset contains audio data from ten cities, provided with a training/test split. The amount of audio in the development dataset is 40 hours. The evaluation set contains a total of 30 hours of audio data, recorded in 12 cities (two cities not encountered in training). 
Evaluation was be performed similarly to Subtask A, using average accuracy across the three acoustic scene classes,  and multi-class cross-entropy. The systems were ranked by the average accuracy.

\vspace{-6pt}

\subsection{System complexity requirements}

This task imposed a classifier complexity expressed in terms of model size on disk. The chosen limit was 500~KB for the non-zero parameters, which means 128~K parameters in the 32-bit float data type (128000 parameters * 32 bits per parameter / 8 bits per byte = 512000 bytes = 500~KB). This approach for limiting the model size allows participants some flexibility in design, for example minimizing the number of non-zero parameters of the network in order to comply with this size limit (sparsity), or quantization of model parameters in order to use lower number of bits. 

The computational complexity of the feature extraction stage is not included in the system complexity estimation. Even though feature extraction is an integral part of the system complexity, there is no established method for estimating and comparing complexity of different feature extraction implementations, therefore we do not take it into consideration, in order to keep the complexity estimation straightforward across approaches. Some special situations for feature representations apply. 
Some implementations may use a feature extraction layer as the first layer in the neural network~- in this case the limit is applied only to the following layers, in order to exclude the feature calculation as if it were a separate processing block.
However, in case of embeddings (e.g. VGGish~\cite{Hershey2017}, OpenL3~\cite{Cramer_2019_ICASSP} or EdgeL3~\cite{Kumari_2019_IPDPSW}), the network used to generate the embeddings counts in the number of parameters.
Additionally, batch normalization layers are skipped from the model size calculation. 





\subsection{Baseline system results}

The baseline system for the task implements a convolutional neural network (CNN) based approach, similar to the DCASE 2019 Task 1 baseline \cite{mesaros_2019_DCASE}. It uses 40 log mel-band energies, calculated with an analysis frame of 40 ms and 50\% hop size, to create an input shape of $40\times500$ for each 10 second audio file. The neural network consists of two CNN layers and one fully connected layer, followed by the softmax output layer. Learning is performed for 200 epochs with a batch size of 16, using Adam optimizer and a learning rate of 0.001. Model selection and performance calculation are done similar to the Subtask A system. 
The model size of the system is 450 KB. For comparison, Table \ref{tab:model-size} presents the model size of the baseline system from Subtask A, together with the performance of both systems on Subtask B.


\section{Challenge results}
\label{sec:results}

This section presents the challenge results and a short analysis of the submitted systems.
The task was very popular in this edition too, with 92 systems from 28 teams submitted to Subtask A and 86 systems from 30 teams submitted to Subtask B. This edition had the largest number of participants to the Acoustic Scene Classification among all editions, with a very balanced interest in both problems. The challenge website contains full details on the systems performance and characteristics \footnote{http://dcase.community/challenge2020/task-acoustic-scene-classification}.


\subsection{Subtask A results and analysis}

\begin{table*}[]
\centering
\caption{Selected systems submitted for Subtask B. Top 10 teams, best system per team.}
\label{tab:subtaskB-top10}
\begin{tabular}{ll|cc|ccc|l}
\toprule
{\textbf{Team}} & {\textbf{\#}} & {\textbf{Accuracy}} & {\textbf{Log loss}}  & {\textbf{Size}} & {\textbf{Param}} & {\textbf{Weights}} & {\textbf{Notes}}\\
\midrule
Koutini\_CPJKU\_2    & 1 & {96.5 \%}    & {0.10}      & {483.5 KB} & {345k} & float16 & pruning, post-training quantization \\
Hu\_GT\_3            & 3 & {96.0 \%}    & {0.12}      & {490.0 KB} & {122k} & int8 & post-training quantization \\
McDonnell\_USA\_3    & 4 & {95.9 \%}    & {0.11}      & {486.7 KB} & {3M} & 1-bit & \\
Suh\_ETRI\_3         & 11 & {95.1 \%}    & {0.27}      & {413.0 KB} & {207k} & float16 & sparse connectivity models, ensemble \\
Chang\_QTI\_1        & 12 & {95.0 \%}    & {0.22}      & {491.2 KB} & {601k} & float16 & pruning, weight sharing across layers \\
Wu\_CUHK\_4          & 14 & {94.9 \%}    & {0.21}      & {299.3 KB} & {153k} & float16 & depth-wise  separable CNN \\
Lee\_CAU\_2          & 23 & {93.9 \%}    & {0.15}      & {494.2 KB} & {126k} & float32 & slim model \\
Naranjo-Alcazar\_Vfy\_1 & 24 & {93.6 \%} & {0.20}      & {496.3 KB} & {127k} & float32 & slim model \\
Kwiatkowska\_SRPOL\_2 & 27 & {93.5 \%}    & {0.16}     & {421.0 KB} & {107k} & float32 & depth-wise separable CNN, ensemble \\
Yang\_UESTC\_3        & 26 & {93.5 \%}    & {0.22}     & {258.0 KB} & {119k} & float16 & slim model \\
\midrule
{Baseline}           &   & {89.5 \%}    & {0.40}      & {450.1 KB} & {115k} & float32 & slim model \\
\bottomrule
\end{tabular}
\end{table*}

The highest accuracy among the 92 systems submitted for this subtask was 76.5\%, for the system of Suh\_ETRI \cite{Sangwon2020}, compared to the baseline system at only 51.4\%. 
Table \ref{tab:subtaskA-top10} presents few details for the best system of the top 10 ranked teams. The official system rank is presented along with a breakdown of accuracy according to device categories. All systems in Table \ref{tab:subtaskA-top10} except Jie\_Maxvision\_1 are ensembles of classifiers. 
Most systems were based on mel energies as feature representation (65 of all submissions)
, in some cases combined with CQT, Gammatone, HPSS, and even with MFCC. All except one system were using deep learning architectures based on e.g. ResNet and dilated convolution, and 51 of the systems report employing ensembles. 
A very large proportion of the submitted systems use at least mixup~\cite{Zhang2018} as data augmentation method (71 of all submissions); many systems used multiple forms of data augmentation including resizing and cropping, spectrum correction, pitch shifting, and SpecAugment, which seems to compensate for the device mismatch. Only a  handful of systems using explicit domain adaptation.   


All systems have significantly higher performance on device A, to which the large majority of training data belongs. Accuracy on devices seen in the training (A, B, C, S1, S2, S3) is also somewhat larger than on unseen ones (D, S7, S8, S9, S10), but the difference is not as large as observed in the case of the baseline system. The generalization properties of the systems are good, with small difference in performance between the development and evaluation datasets. 



Log-loss shows the uncertainty of the classifier, and it is lowest among all 90 systems (0.755) for Liu\_UESTC \cite{Liu2020a}, but not for their best system in terms of accuracy (therefore not the system in Table \ref{tab:subtaskA-top10}).  From the systems in Table \ref{tab:subtaskA-top10},  Hu\_GT \cite{Hu2020} has much smaller log-loss than the top system, indicating that the top system is more uncertain of its decisions, even though they have the same accuracy (accuracies of the two systems are within the 95\% confidence interval).



\subsection{Subtask B results and analysis}

The highest accuracy among the 86 submitted systems for this subtask was 96.5\%, compared to an accuracy of 89.5\% for the baseline system. Here, the core task was to find low-complexity solutions for a classification problem that was not too difficult. In terms of log-loss, the systems in this subtask are more confident in their decisions than the ones in Subtask A, with the best log-loss value belonging to the top system. 

Details about performance and system choices for the best system of the top 10 ranked teams are presented in Table~\ref{tab:subtaskB-top10}.
Many of the submissions used deep learning architectures and imposed restrictions on the models and their representations, such as using slim models, depth-wise separable CNNs, pruning and post-training quantization of model weights. 
The top system Koutini\_CPJKU~\cite{Koutini2020} used a combination of pruning and quantization, using 16-bit float representation for the model weights and having a reported sparsity of 0.28 (ratio of zero-valued parameters). A much sparser model was obtained by Chang\_QTI~\cite{Chang2020} through pruning and weight sharing across layers, while many participants used quantization and 16-bit float representations. 




One  solution very different from the others was presented by McDonnell\_USA \cite{McDonnell2020}, that used a single bit per weight, and represented the model weights as $-1$ or $+1$. This kind of bit-wise neural network can be seen as an extreme form of quantization which allows using a very large model: its reported parameter count is 3M, compared to an average of few hundreds of thousands for the other systems. 
Systems that used the default 32 bit float representations opted for models specifically designed to fit in the required limit (named "slim" models in Table~\ref{tab:subtaskB-top10})~\cite{Naranjo-Alcazar2020_t1} or depth-wise separable CNNs.
It is notable that most systems in Table \ref{tab:subtaskB-top10} have models close to the maximum allowed size, except two, one that uses depth-wise separable CNN and another one using a slim model. 
Overall, the size of the models ranges from 8.8KB (ranked 82) to 499.5 KB (ranked 53), with 47 systems having a model size over 400 KB. In terms of number of parameters, the system by McDonnell\_USA \cite{McDonnell2020} is an outlier with 3M, with next largest number of parameters being 600 k. 







%

\section{Conclusions and future work}\label{sec:conclusions}

This paper presented the results of the DCASE 2020 Challenge Task~1 Acoustic Scene Classification. The two different subtasks offered in this edition were oriented towards real-world applications, in one case robustness and generalization to multiple devices, and in the other case the need for low-complexity solutions. A number of  solutions were presented, with data augmentation being the main tool used for the generalization problem, while the majority of the low-complexity solution were based on quantization of model parameters, with the top systems in each subtask using a combination of techniques. Considering the high number of submissions, the task was very successful. In particular, the low-complexity requirement attracted a very high number of participants, indicating great interest for the topic in the academic environment in addition to industry. For the future editions, we hope to be able to provide a more accurate and universal approach to measure model or computational complexity, in order to establish a widely accepted methodology for measuring characteristics of methods intended for limited-resource devices.

\vspace{-6pt}
\section{ACKNOWLEDGMENT}
\label{sec:ack}
This work was supported in part by the European Research Council under the European Unions H2020 Framework Programme through ERC Grant Agreement 637422 EVERYSOUND.

\bibliographystyle{IEEEtran}
\bibliography{refs}

\end{sloppy}
\end{document}